# A picture of pseudogap phase related to charge fluxes

Xi Chen[1,†], Jiahao Dong[1,†], Xin Li[1,*]

1. John A. Paulson School of Engineering and Applied Sciences, Harvard University

29 Oxford St, Cambridge, MA 02138

**Abstract**

Recently, charge density fluctuations or charge fluxes attract strong interests in understanding the unconventional superconductivity. In this paper, a new emergent configuration in cuprates is identified by density functional theory simulations, called the charge pseudoplane, which exhibits the property of confining the dynamic charge fluxes for higher superconducting transition temperatures. It further redefines the fundamental collective excitation in cuprates as pQon with the momentum-dependent and ultrafast localization-delocalization duality. It is shown that both pseudogap and superconducting phases can be born from and intertwined through the charge flux confinement property of the charge pseudoplane region. Our experimental simulations based on the new picture provide good agreements with previous angle resolved photoemission spectroscopy and scanning tunneling microscopy results. Our work thus opens a new perspective on the origin of pseudogap phase and other related phases in cuprates, and further provides a critical descriptor to search and design higher temperature superconductors.

*: Corresponding author: lixin@seas.harvard.edu

†: Equal contribution

**Introduction**

Understanding high temperature superconductivity has caused a long exciting debate in the community of condensed matter physics since the discovery of $La_2CuO_4$ family of superconductors[1]. Earlier works identified the $CuO_2$ plane as the superconducting plane, where strong electronic correlations and spin fluctuations exist[2–4]. Although all cuprate superconductors have the same $CuO_2$ building blocks, previous works have pointed out the importance of structural and electronic features outside the $CuO_2$ plane in modulating the material dependence of the superconducting transition temperature $T_c$[5–13]. However, recent reports of the light-induced superconducting-like behavior near room temperature in $YBa_2Cu_3O_{6+x}$ (YBCO)[14–16] and the femtosecond electron-phonon lock-in phenomenon in iron selenide[17] further suggest the importance of ultrafast dynamical processes in unconventional superconductors, which many previous models neglect.

In our recent model[18,19], ultrafast lattice effect was explicitly considered to couple with transient electronic effects. Specifically, higher apical lattice oscillation frequency and stronger apical charge flux were correlated with the maximum superconducting transition temperature $T_{c,max}$ among different cuprate families. Modulated by such dynamical effects, a spatially localized hole can hop from one Cu site to the neighboring one with vanishingly small energy barrier. In this paper, we show that a fundamental collective excitation in cuprates can emerge from the collective behavior of charge fluxes to help elucidate the origin of the pseudogap phase, which is a central puzzle of high-temperature superconductivity[20]. For a given underdoped cuprate, the state above $T_c$ but below a characteristic temperature range of $T^*$, which can be determined from many experimental techniques, features a spectral gap of mysterious nature, labelled as the pseudogap state or pseudogap phase. Its classification as a true thermodynamic phase, associated with broken long-range symmetries, however, is still a subject of debate[21–30]. Recent photodestruction experiment suggests, on an ultrafast timescale, the existence of ultrafast carrier localization into polaronic states[31]. On the other hand, in angle resolved photoemission spectroscopy (ARPES) measurements, pseudogap phase is known to exhibit a more quasiparticle-like behavior around the nodal direction[32], as with doping, relatively sharp peaks appear near the Fermi level, giving the Fermi arc; while simultaneously around the antinodal direction such sharp peaks are completely suppressed [33], defining the pseudogap. Meanwhile, along both nodal and antinodal directions a broad distribution of spectrum weight was observed to extend far below the Fermi level. To explain the complicated experimental results in the pseudogap phase, the polaron-like behavior of charge carriers[34,35], the spin or charge fluctuation contributions[36–39] and the local pair models [40–44] were proposed. There were thus also debates about the degree of localizations for the holes in the pseudogap phase of cuprates.

In this paper, we first introduce the concept of "charge pseudoplane region" that governs the collective flows of charge fluxes in cuprates, through which more localized and more delocalized states of holes can dynamically transform into each other on an ultrafast timescale with a momentum dependence. This correspondingly redefines a new fundamental collective excitation in cuprates with the momentum-dependent and ultrafast localization-delocalization duality, based upon which simulated ARPES spectra agree well with previous ARPES experiments in a broad doping range, including the pseudogap phase region. More importantly, a comparison with experimental measurements on pseudogap boundary temperature $T^*$ versus doping shows that under this picture the maximum strength of such duality interaction, which is governed by the charge flux confinement property of the pseudoplane region, is strongly correlated with $T_{c,max}$ among different cuprate families, suggesting that both pseudogap and superconducting phases are intertwined through such flux confinement property. Furthermore, a percolation simulation based on this picture at various doping levels also gives an interpretation of the temperature independent critical doping around 19% reported recently[45], and a connection to previous scanning tunneling imaging results on inhomogeneity[46–49]. Our work thus reconciles previous results of the pseudogap phase from various experimental techniques with different momentum, spatial and time resolutions, and provides a new route to design and search higher temperature superconductors.

## Results and Discussions

### Dynamical Charge-Flux-Confining Pseudoplanes

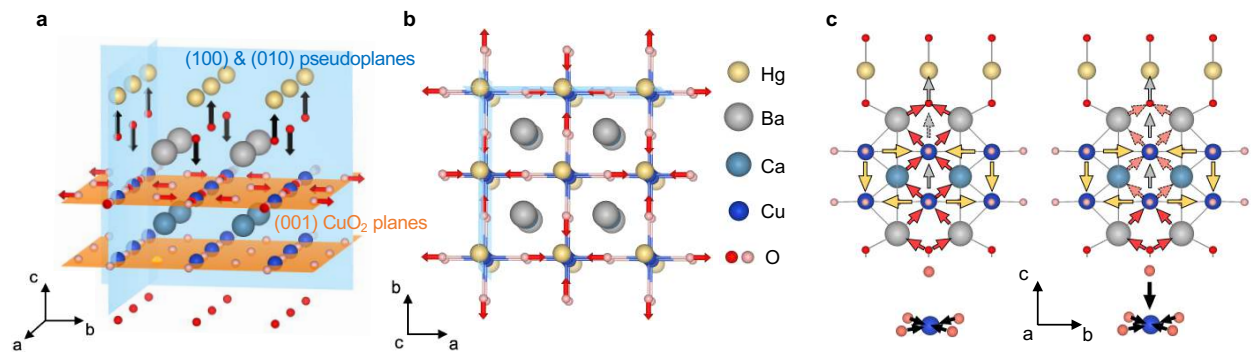

**Figure 1. Emergent charge-confining pseudoplane structures. (a)** pseudoplane structures (blue planes) in HgBa$_2$CaCu$_2$O$_6$ (Hg-1212). The black and red arrows indicate the movement of oxygens in the apical mode and breathing mode, respectively. **(b)** Same structure viewing along the $c$ direction. **(c)** Charge fluxes induced by the oxygen breathing mode (left panel), and the breathing mode coupled to the apical mode (right panel). Solid red arrows indicate stronger leakage fluxes than dashed pink ones. Solid grey arrows indicate stronger vertical fluxes than the dashed grey arrows. The ion distortion patterns of the corresponding oxygen modes are shown in the bottom.

Figure 1 shows the (100) and (010) atomic planes that are perpendicular to the CuO$_2$ planes. To distinguish such vertical plane from the CuO$_2$ plane, it is called the pseudoplane hereafter. "Pseudo" here refers to two facts that will be discussed in details below: first, its novel electronic property relies on the definition of a thickness (such as 1 Å) for the "plane" region that contains certain electronic orbitals of the included atoms; second, such novel property governs the pseudogap phase. Each pseudoplane cuts through the apical and the in-(CuO$_2$)-plane (in-plane hereafter) oxygen ions, the in-plane copper ions and the apical atoms (Hg, Tl, etc.), which contains half lobes of the $d_{x^2-y^2}$ orbitals of the in-plane Cu ions, the $d_{z^2}$ orbitals, and also the $s$ or $p$ orbitals of the apical atoms. Note that these orbitals are also hybridized with oxygen $p_x$, $p_y$ and $p_z$ orbitals. With phonon perturbations, dynamic charge fluxes are generated in DFT simulations (Fig 1c, see Methods). Fig. 1c presents the strength and direction of charge fluxes induced by the breathing oxygen mode (left panel), compared with the fluxes induced by the breathing mode coupled to the apical oxygen mode (right panel), using Hg-1212 as the example (see Methods and Supplementary Fig. S1). When the two oscillation modes are coupled, we notice that the charge fluxes are more confined within the pseudoplane (grey and yellow arrows). Especially the fluxes along the vertical in-pseudoplane pathway, from Cu to apical oxygen further to Hg, are enhanced. Meanwhile, the fluxes along the out-of-pseudoplane pathways, from Cu to Ba further to apical oxygen, and from Cu to Ca further to the Cu in the neighboring $CuO_2$ layer (red and pink arrows), are reduced (see Supplementary Fig. S2).

The most intriguing property of these pseudoplanes is that they as an integrity form the pseudoplane region (i.e., all the parallel and intersected pseudoplanes in a cuprate crystal), which dynamically confines the charge flux oscillations inside the region, by driving those phonon couplings that prefer such charge flux confinement. In order to quantify such flux confinement property that is a dynamic collective behavior of all the relevant electron orbitals, we define a thickness layer of 1 Å for each of these pseudoplanes. Electron densities within such layers are defined as being inside the pseudoplane region, while electrons in other regions in the cuprate structure are considered as being outside the pseudoplane region. Note that, however, our conclusions do not rely on the thickness to be exactly 1 Å. Using this metric, we further calculate different levels of extra charge flux leakage outside the pseudoplane region induced by all pairwise couplings of the main phonon modes in a supercell (Fig. 2a and 2b, also see Methods), where results of $HgBa_2CaCu_2O_6$ (Hg-1212), $HgBa_2CuO_4$ (Hg-1201), YBCO, and LSCO are compared.

Some important trends are observed in Fig. 2. First, phonon modes that are more strongly coupled, represented by regions with darker pixels in Fig. 2a, tend to show stronger charge confinement ability, or less extra charge leakage, corresponding to regions with more blueish pixels in Fig. 2b. This reflects the charge flux confinement property of pseudoplanes to drive the coupling of those phonon modes that can better confine the charge fluxes within the pseudoplane region. Second, such phenomena show a family dependence for cuprate superconductors. YBCO ($T_{c,max}$ = 93 K) and Hg-1201 ($T_{c,max}$ = 94 K), with lower $T_{c,max}$ than Hg-1212 ($T_{c,max}$ = 127 K), show a weaker such correlation in general, while LSCO ($T_{c,max}$ = 38 K) with the lowest $T_{c,max}$ among these families shows the weakest correlation. Specifically, for Hg-1212 there are three frozen modes (#19-21), corresponding to the apical oxygen mode, the breathing and the anti-breathing oxygen modes, respectively, with obvious couplings to many perturbation modes. They show the strongest phonon coupling and the best charge confinement ability to lock oscillating charge fluxes inside the pseudoplane region. This relation also applies to the other cuprate families, where the bottom three modes (Hg-1201) or four modes (YBCO and LSCO) include the apical, and breathing-like modes.

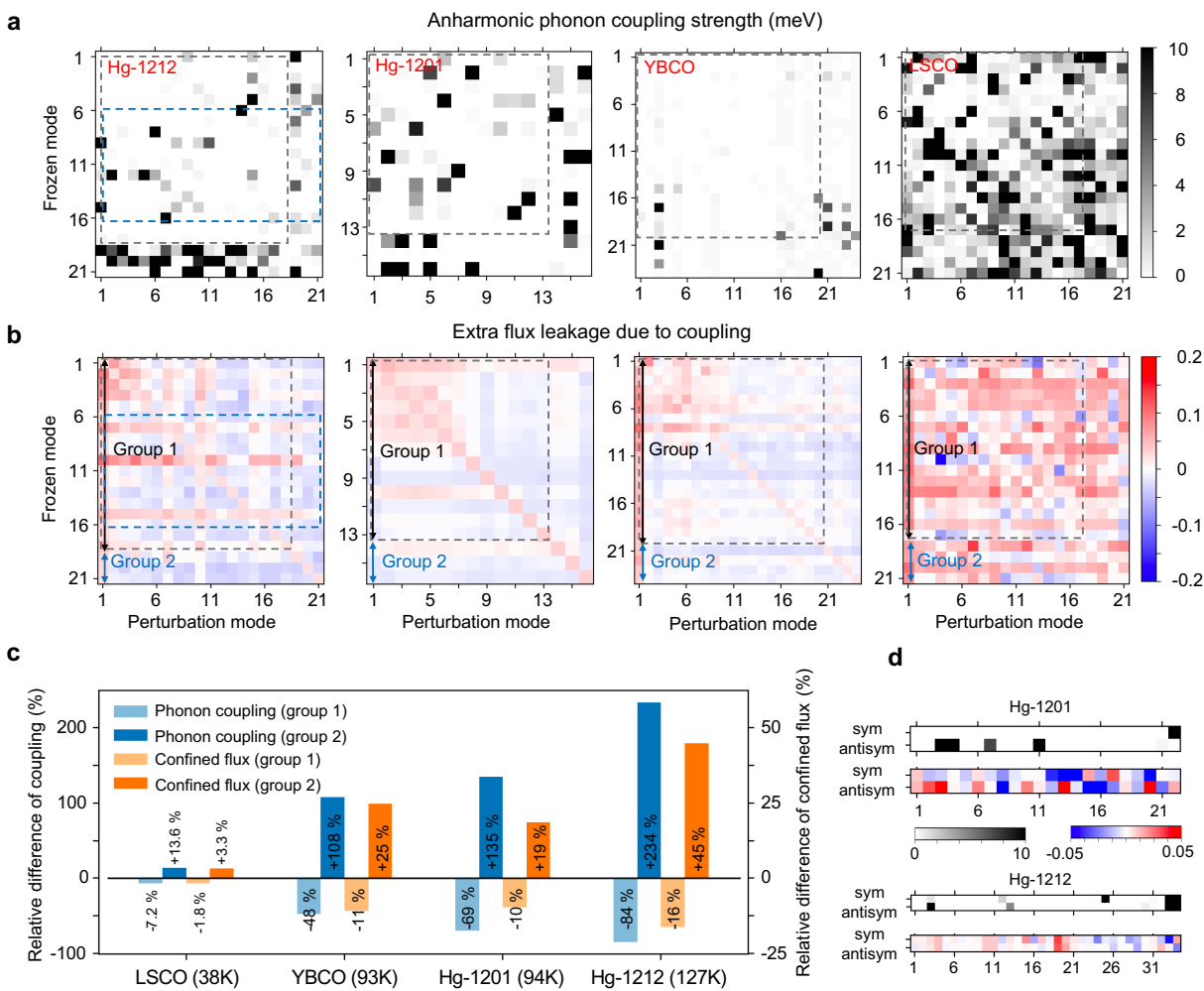

**Figure 2. Interplay between phonon coupling and charge flux confinement through the pseudoplane region for higher $T_{c,\mathrm{max}}$.** **(a)** Anharmonic phonon coupling strength of 21, 16, 24 and 21 modes of Hg-1212, Hg-1201, YBCO and LSCO, respectively. Grey scale intensity is proportional to the coupling strength. **(b)** Extra flux leakage out of the pseudoplane region due to the phonon coupling. The extra flux leakage here is defined by the ratio $f_o/(f_o+f_i)$, where $f_o$ and $f_i$ are the extra charge flux to atoms outside and inside the pseudoplane region, respectively, due to perturbations from each pair of phonons. The result is already with the average value of the entire map subtracted. The blue or red color of each pixel hence indicates decreased or increased leakage by each pair compared with the mean extra flux leakage value

induced by all phonon pairs (See Methods). The phonon pairs are separated into two groups, where each pair in group 2 (the pixels outside the gray dashed square) consists of at least one oxygen breathing-like mode or apical-like mode, while all the pairs in group 1 (the pixels inside the gray dashed square) do not involve these modes. The blue dashed box in the Hg-1212 case in 2a, 2b (frozen mode #6-16) includes phonon modes that involve oxygen rotations as illustrated in Supplementary Fig. S4, and $B_{1g}$-like bucklings (mode #9, #12). **(c)** Correlations between the anharmonic phonon coupling strength and the flux confinement ability for LSCO, YBCO, Hg-1201 and Hg-1212. Lighter or darker blue bars show the relative difference between the average phonon coupling strength of group 1 or group 2 and the average of all mode pairs, respectively. Orange bars show the corresponding relative differences of the flux confinement ability for group 1 and group 2. **(d)** Anharmonic phonon coupling (grey scale) comparison between the symmetric/antisymmetric apical oxygen modes and all other phonon modes in the phonon spectra for Hg-1201 and Hg-1212, and the corresponding extra flux leakage (color scale). The apical modes are set as the frozen mode and other modes as the perturbation mode. Here the flux leakage is calculated based on the generalized pseudoplane of 1 Å thickness $CuO_2$ slab, and without the mean value subtracted.

To more quantitively illustrate the correlation between the flux confinement ability of the pseudoplane region and the $T_{c,max}$ of different cuprate families, we separate the phonon pairs to group 1 and group 2, where mode pairs in group 2 involve the apical or breathing-like modes, while that in group 1 do not. We calculate the average phonon coupling and flux confinement ability for mode pairs in the two groups, relative to the average values of all mode pairs (Fig. 2c). For all four materials of LSCO, YBCO, Hg-1201 and Hg-1212, mode pairs in group 2, i.e., those involving the apical or breathing-like modes, show stronger phonon couplings on average (positive values) than the ones in group 1 (negative values). Similarly, mode pairs in group 2 also show a better flux confinement ability. More importantly, this metric shows a correlation to the material dependence of $T_{c,max}$, becoming stronger from LSCO ($T_{c,max}$ = 38 K) to YBCO ($T_{c,max}$ = 93 K), Hg-1201 ($T_{c,max}$ = 94 K) and Hg-1212 ($T_{c,max}$ = 127 K). Therefore, the ability of charge pseudoplanes to confine the charge fluxes shows certain positive correlation with the experimental $T_{c,max}$ of different hole doped cuprate families. This metric is related to the apical force metric[18], where stronger apical flux was generated by the perturbation and associated with higher $T_{c,max}$. However, the flux confinement ability is a more sophisticated metric, which we will show is critical to the understanding of the pseudogap phase. Note that the frozen modes #9 and #12 in group 1 of Hg-1212 involve the $B_{1g}$-like components of buckling modes. They couple with more perturbation modes to confine the charge fluxes than other modes, but less than the bottom three modes of #19-21. In addition, at the particular k point of (0.5, 0.5, 0) used to generate the phonon spectra in Fig. 2, some other modes of interest, such as half-breathing and $A_{1g}$ modes, are not allowed. Therefore, our selection of group 2 modes here only focuses on the ones with the strongest correlation between phonon coupling and flux confinement to illustrate the importance of the new metric, rather than serving as a complete classification of phonon modes based on such metric.

The coupling of apical and breathing modes was also argued to be crucial to the transport property of cuprates, while the symmetric and antisymmetric apical modes exhibit a $CuO_2$ layer dependent effect on the hole localization [18]. In Fig. 2d, we show that for Hg-1201 with a single $CuO_2$ layer the symmetric and antisymmetric apical modes show very different couplings with other modes in the phonon spectrum. Similarly, the extra flux leakages induced by these couplings are also different between the symmetric and antisymmetric cases. While for Hg-1212 with double $CuO_2$ layers, the effect due to the apical mode symmetry is greatly reduced. Note that in this particular calculation we focused on the horizontal $CuO_2$ plane as a generalized pseudoplane, rather than the vertical pseudoplanes, since the $CuO_2$ plane is directly and strongly affected by the apical mode symmetry. The calculations in Fig. 2c, 2d thus also suggest that the generalized flux confinement ability may be less perturbed by apical mode symmetries, and hence enhances with increasing $CuO_2$ layers. We further replace the apical atom Hg in Hg-1212 with Tl, Bi, Y, and Pb and calculate the phonon coupling of these hypothetical materials as we did in Fig. 2a. SI Fig. S3 shows that the total phonon coupling strength of Hg-1212 is higher than the hypothetical structures, suggesting that previous experimentalists might have identified one of the best apical atoms, i.e., Hg, for high $T_c$ in this family.

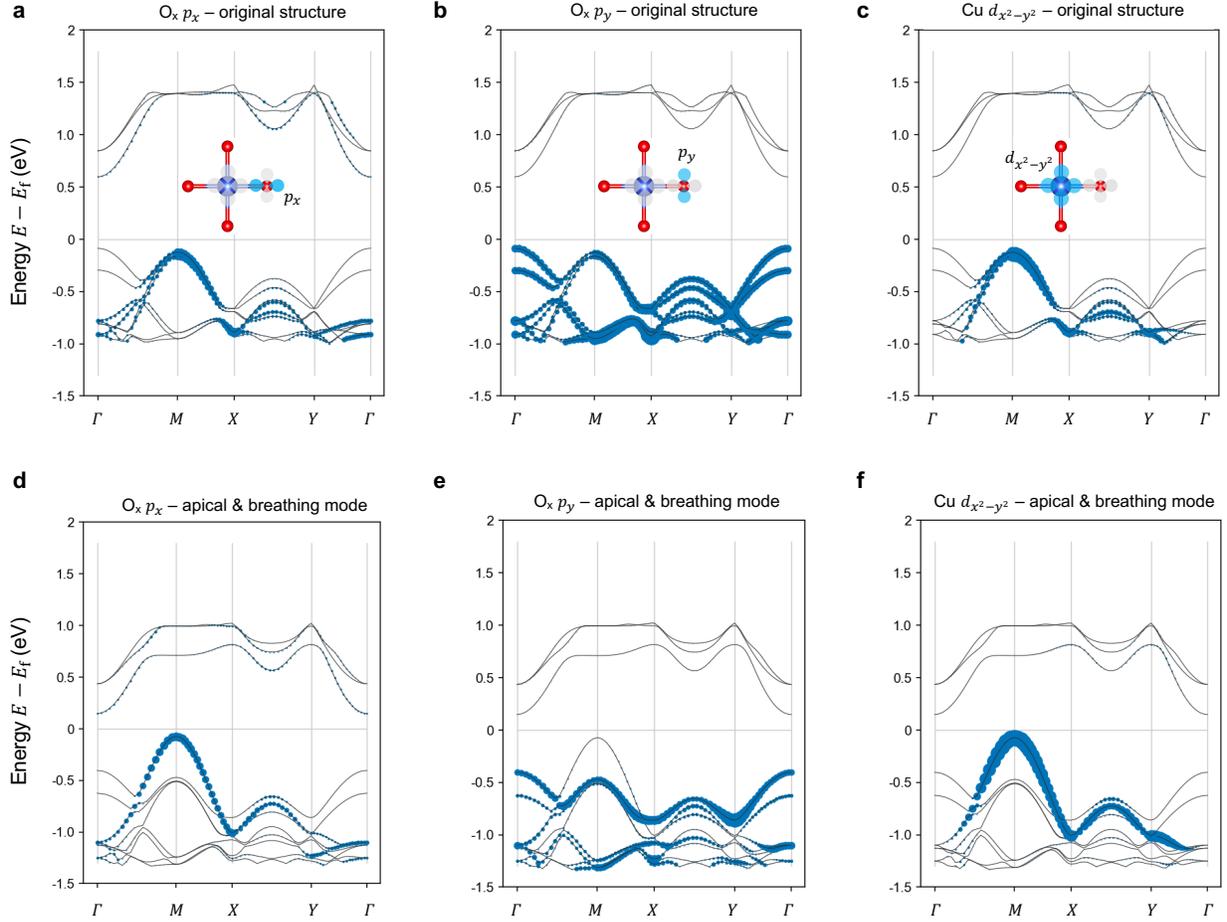

**Figure 3. Effect of phonon coupling on the band occupation of $p_x$ and $p_y$ orbitals of oxygens along the $x$ bonding direction ($O_x$) and Cu $d_{x^2-y^2}$ orbitals**. **(a-f)** Comparison between the original electronic structures without phonon distortions **(a-c)** and that with coupled apical and breathing oxygen modes **(d-f)**, calculated using Hg-1212. The black curves are the calculated band structure and the size of blue spheres indicate the orbital occupation in each band at each k point. The vertical gray lines mark the high symmetry points and the horizontal gray line marks the Fermi level ($E_f$). The coordinates of the high symmetry points are: Γ (0,0,0), M (0.5,0.5,0), X (0.5,0,0), Y (0,0.5,0).

The dynamic charge flux confinement property modulated by the pseudoplane region, as we analyzed in the real space in Figure 2, in fact reflects the collective dynamic property of electronic structures in the momentum space. We illustrate this connection between real and momentum spaces using the coupled breathing and apical modes, as it is one of the couplings that well confine the charge flux flows within the pseudoplane region. Figure 3 shows the band occupation of the $p_x$ and $p_y$ orbitals of oxygens along the x bonding direction ($O_x$), and the copper $d_{x^2-y^2}$ orbital around Fermi level ($E_f$), compared between the original structure without phonon oscillation and the perturbed structure with coupled breathing and apical modes (See Fig. 3a-c insets for illustration of the orbitals). The most distinct difference in the band occupation exists around the M high-symmetry point with the wavevector (0.5,0.5,0), which is the momentum along the nodal direction with the largest tendency of extra charge flux leakage out of the pseudoplane region. Compared with the original unperturbed structure, where all the three orbitals (Fig. 3a-c) share the occupation of the highest band below $E_f$ around the M point, the structure with coupled breathing and apical modes (Fig. 3d-f) show a complete suppression of the $p_y$ orbital occupation around the M point in the highest band below $E_f$, while the occupation there from the $p_x$ and $d_{x^2-y^2}$ orbitals are enhanced. Note that the $O_x$ $p_x$ and Cu $d_{x^2-y^2}$ orbitals (Fig. 3a, 3c insets) are hybridized inside the pseudoplane region, their enhanced occupation contributes to a better charge flux confinement inside the pseudoplane region. Meanwhile, the suppression of the $p_y$ orbital of $O_x$ oxygen that points toward the out-of-pseudoplane direction (Fig. 3b inset) corresponds to a reduced flux leakage from the pseudoplane. On the contrary, for the oxygen rotation modes (Supplementary Fig. S4) that in general induce a weak phonon coupling and strong charge leakage (Fig. 2ab), the ion perturbation induced change in the band occupation is opposite. The occupation of the highest band below $E_f$ by the $O_x$ $p_y$ orbital around the M point is even increased (Supplementary Fig. S5) compared with the original structure. More electronic structure details corresponding to the charge flux confinement property related to the coupled breathing and apical modes are further shown in Supplementary Fig. S6 and S7. Supplementary Figure S6 shows that compared with just the breathing mode, adding the coupled apical mode increases the band occupation of the $O_x$ $p_x$ and Cu $d_{x^2-y^2}$ orbitals at X and Y symmetry points for a better in-pseudoplane flux confinement. The orbitals from other atoms of Hg-1212 also exhibit a similar trend. Considering Ba that locates out of the pseudoplane region, the flux onto Ba corresponds to the out-of-pseudoplane flux leakage. Supplementary Fig. S7 shows a similar reduction of band occupation around M point by Ba orbitals, and hence an enhanced charge flux confinement, with coupled apical and breathing modes; while an increase of band occupation for charge leakage is observed with coupled rotation and apical modes. In summary, these observations in real and momentum spaces suggest that not only stronger charge fluxes[18], but more fundamentally as we will discuss, a better confinement of their collective oscillations by the charge pseudoplane region may prefer higher $T_{c,max}$ for a cuprate family.

**Collective Excitation with Momentum-dependent and Ultrafast Localization-delocalization Duality**

Our vision is that the dynamic charge flux confinement property of the pseudoplane region plays a critical role here in distinguishing the hole representations. Because as long as the hole moves, a group of charge fluxes is generated, which will be modulated by the charge pseudoplanes. Generally speaking, a hole moving along the antinodal direction (*x* or *y* direction) won't obviously violate such charge flux confinement property of the pseudoplane, as at least the direction of charge movement associated with the direct hole hopping is inside the pseudoplane (Fig. 1, Fig. 4a); while moving along the nodal direction (45º from *x* or *y* direction) generates an immediate tendency for charge fluxes to leak outside the pseudoplane region. In order to confine these fluxes, a delocalization into weaker fluxes with decomposed momentums along antinodal directions inside the pseudoplane region, i.e., a flux decomposition process to better confine the fluxes of a hole moving along the nodal direction, will be preferred. Under this picture, previous ARPES results may actually suggest that the hole movement along the antinodal direction is well protected by the charge pseudoplane to behave more localized in transport, while once hopping along the nodal direction, it may show a delocalization to confine fluxes through the flux decomposition process and some flux leakage due to the lack of perfect pseudoplane protection. Such a picture is consistent with our DFT simulations (Fig. 4b), where a localized hole hopping along the nodal direction tends to be delocalized at the early stage of conversion due to the lack of an energy barrier, while the antinodal hopping well protects the spatially localized state by the immediate energy barrier. It is worth noting that in the later stage of conversion beyond 30% along the nodal direction, an energy barrier gradually builds up to prevent the further delocalization. That is, even along the nodal direction the delocalization of hole is still limited in space and may still be transiently re-localized due to the fact that in the simulation (and in practice), no exact diagonal hopping can happen without any perturbative components of lattice oscillations and flux flows.

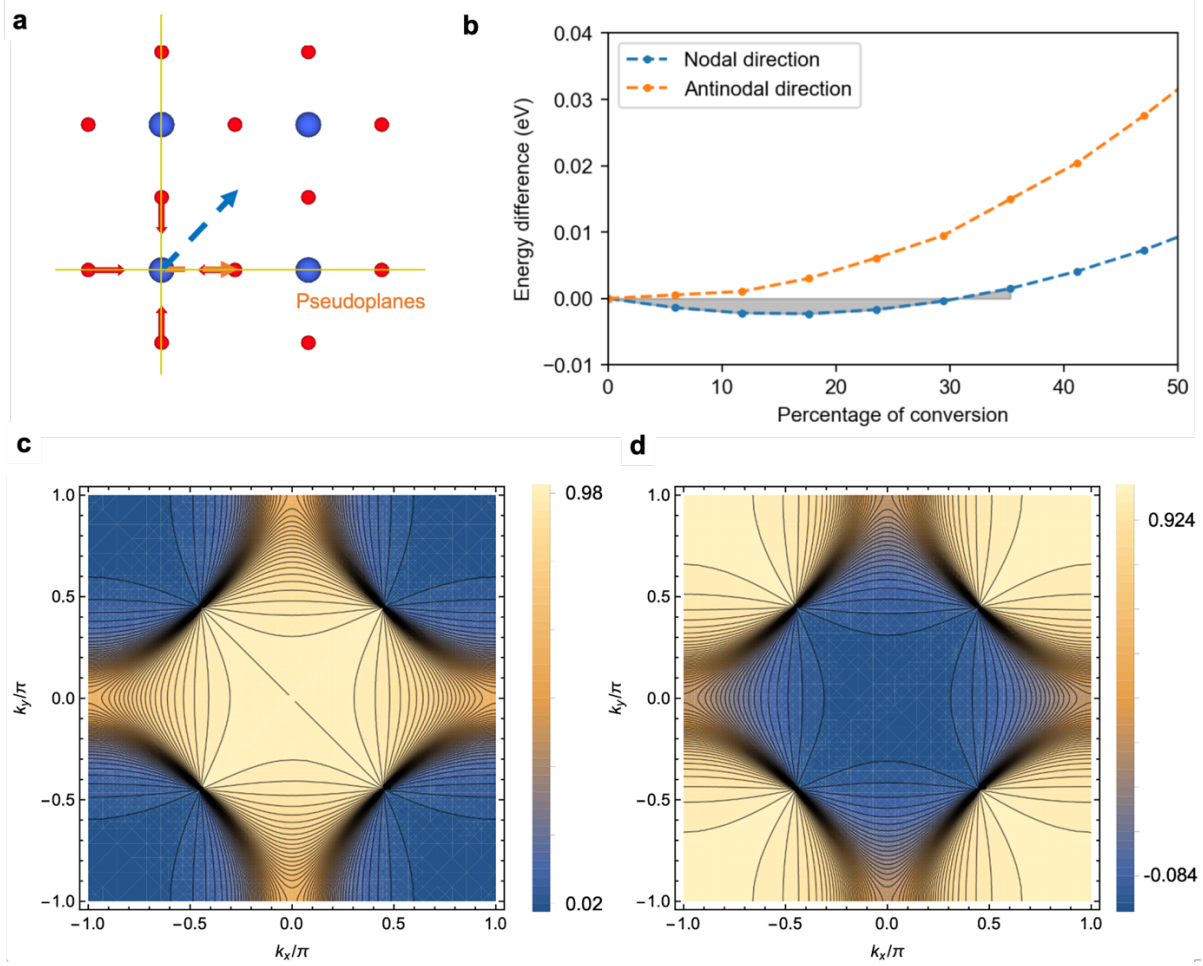

**Figure 4. Momentum-dependent localization-delocalization duality of pQon. (a)** Illustration of a localized hole at a Cu site, and its nodal (blue arrow) and antinodal (orange arrow) hopping directions. **(b)** DFT calculated energy curves of converting the localized hole at 0% of conversion illustrated in (a) toward a completely delocalized quasiparticle at 100% of conversion along the nodal and antinodal hopping directions, respectively. **(c)** The percentage of delocalized state $c_{k,\sigma}^{\dagger}$ from $|m_{k,\sigma}|/(|m_{k,\sigma}| + |n_{k,\sigma}|)$, **(d)** The percentage of localized state $d_{k,\sigma}^{\dagger}$ from $|n_{k,\sigma}|/(|m_{k,\sigma}| + |n_{k,\sigma}|)$. T = 100K and p = 0.08 for the calculations of (c) and (d) based on Eq. (4) and (5).

We now formularize this picture in cuprates and show that a new fundamental collective excitation can be defined in such systems. We use $c_{k,\sigma}^+$ and $d_{k,\sigma}^+$ to represent creation operators of spatially more delocalized and spatially more localized carrier states, respectively, and propose the following Hamiltonian:

$$H = \sum_{k,\sigma}(E_k c_{k,\sigma}^\dagger c_{k,\sigma} + E_d d_{k,\sigma}^\dagger d_{k,\sigma} + U_f(t) d_{k,\sigma}^\dagger c_{k,\sigma} + U_f^*(t) c_{k,\sigma}^\dagger d_{k,\sigma}) \quad (1)$$

Where $E_k$ and $E_d$ are bare kinetic energies of the more delocalized and more localized states, respectively, i.e., kinetic energies when they propagate freely. The interaction term $U_f$ represents the strength of the dynamic charge transfer between the two states of $E_k$ and $E_d$, which are induced by coupling to the lattice and/or electronic correlations as modulated by the charge pseudoplanes. We set the time average of the interaction term between the two states, $<U_f(t)>$, as a $d$-wave type function in Eq. (2) to exhibit the d-wave-like directional preferences as required by the charge pseudoplane region and consistent with the above DFT results, while without explicitly considering its ultrafast time dependence on $t$ in this paper. The time average treatment is technically adequate for the following comparison with ARPES experiments that are also from the time average, however, it is worth noting that $U_f(t)$ is originated from the collective flux flows that are intrinsically ultrafast in nature, which should manifest the importance of the time dependence in other ultrafast experimental techniques.

When a hole moves in the axial or antinodal directions approaching the Brillouin zone boundary, the more delocalized state hybridizes with the more localized state strongly due to the charge flux confinement property of the pseudoplane region, because although transient delocalization through the flux decomposition process is a way to prevent the leakage induced by all kinds of perturbations, it prefers to transform back to the more localized state that better confines fluxes intrinsically when moving along the axial direction. $U_f$ reaching the maximum in the axial directions in Eq. (2) reflects that such transformation between the two states is the strongest approaching the antinodal direction. The more delocalized state hybridizes less with the more localized state when a hole moves in the diagonal or nodal directions approaching the Fermi level, because the more delocalized state now can obviously confine fluxes better than the more localized state through the flux decomposition process. $U_f = 0$ in the exact diagonal direction reflects that the hole in principle tends to stay in the more delocalized state without transforming back to the more localized state, although any perturbations that induce slight deviations from the exact nodal direction can still transiently transform it back to the more localized state. It is thus also expected that hole moving around the diagonal direction shows a higher probability in the more delocalized state approaching the Fermi level. Since the confinement property of charge pseudoplanes represents the collective behavior of the orbitals and their interactions with lattice oscillations inside the pseudoplane region, where the $d$-orbitals of Cu ions are critical, the simplest reasonable assumption of $U_f$ that describes the above picture is a $d$-wave form in Eq. (2), when we assume that the value of $U_f$ evolves continuously between the nodal and antinodal directions.

$$U_f = <U_f(t)> = V_f|cosk_x - cosk_y| \qquad (2)$$

In Eq. (2), $V_f$ measures the maximum strength of the interaction.

Diagonalization of Eq. (1) yields two renormalized bands:

$$\omega_{k,\sigma}^{\pm} = \frac{E_k + E_d}{2} \pm \sqrt{(\frac{E_k - E_d}{2})^2 + |U_f|^2} \qquad (3)$$

For the lower renormalized band $\omega_{k\sigma}^-$, the corresponding operator is:

$$b_{k,\sigma}^{\dagger} = m_{k,\sigma}c_{k,\sigma}^{\dagger} + n_{k,\sigma}d_{k,\sigma}^{\dagger} \qquad (4)$$

where the coefficients have the following expressions:

$$m_{k,\sigma} = \frac{U_f}{\sqrt{(\omega_- - E_k)^2 + |U_f|^2}}, \quad n_{k.\sigma} = \frac{\omega_- - E_k}{\sqrt{(\omega_- - E_k)^2 + |U_f|^2}} \quad (5)$$

The operator $b_{k,\sigma}^\dagger = m_{k,\sigma} c_{k,\sigma}^\dagger + n_{k,\sigma} d_{k,\sigma}^\dagger$ represents a carrier with a dual nature of more delocalized and more localized states. We thus define $b_{k,\sigma}^\dagger$ as the new fundamental collective excitation in this system, coined as pQon, where "p" refers to the more localized property of polaron, while "Q" refers to the more delocalized property of quasiparticle. However, pQon is different from the conventional meaning of either polaron or quasiparticle, as propagating pQon will be in a dynamical ultrafast transformation between more localized and more delocalized states, the ratio of which in the time average is strongly momentum-dependent. When pQon propagates near the diagonal or nodal direction approaching the Fermi level, it stays longer at the more delocalized Q-state, and in time average the coefficient $m_{k,\sigma}$ for the Q-state is much larger than $n_{k,\sigma}$ for the more localized p-state (Fig. 4c, 4d); while near the axial or antinodal direction approaching the Brillouin zone boundary, pQon reflects strong electronic interactions of the pseudoplane region, making it behave more localized in time average throughout the propagation, with coefficient $n_{k,\sigma}$ comparable to $m_{k,\sigma}$ (Fig. 4c, 4d). The transformation between the two states is on the ultrafast femtosecond timescale, with the momentum dependence modulated by the pseudoplane region through the collective confinement behavior of charge fluxes. Note that previous models on the boson-fermion coupling picture [50,51] shows different expressions in Hamiltonian, while the Kondo model describes the behavior of quantum impurities. Our model and picture here instead redefine the fundamental excitation in cuprates as pQon (i.e., to play the role of hole) with the momentum-dependent and ultrafast localization-delocalization duality.

**ARPES Simulations based on pQon**

Inspired by previous pioneering ARPES simulations based on different models[37,52–54], we further perform ARPES simulations based on the electronic structure of the unique pQon model (see Methods). We set $E_d = -0.1E_k$, where the exact value of 0.1 here has no more physical significance than the mobility consideration to represent the much lower mobility of the more localized p-state than the more delocalized Q-state. The negative sign refers to the assumption that the p-state of pQon behaves just hole-like in the band structure near the Fermi level for hole doped cuprates, while the Q-state formed by a cloud of charge fluxes behaves electron-like that easily extends above the Fermi level, analogous to the pair of valence band hole and conduction band electron in semiconductor physics. To further appreciate the physical importance of this assumption, we note that the BCS electronic spectrum considered under the framework of our model is also equivalent to the hybridization of a hole band and an electron band. Because in Eq. (3), if we take $E_d = -E_k$ and $U_f = \Delta_k = \Delta_0$ instead, the eigenvalues become $\omega_{k,\sigma}^{\pm} = \pm\sqrt{E_k^2 + |\Delta_0|^2}$, which recovers the electronic spectrum of BCS superconductors. We also note that the negative sign, as well as the nonzero small value of $E_d$ in $E_d = -0.1E_k$ is critical to reproduce in our simulations the two bending points observed in the ARPES experiment, which will be discussed later. For $E_k$ (and $E_d$), we employ a tight-binding dispersion relation on a square lattice $E_k = -2t_1(cosk_x + cosk_y) - 4t_2 cosk_x cosk_y - \mu$, where $t_1$, $t_2$ and $\mu$ represent nearest- and next-nearest-neighbor hoppings, and chemical potential, respectively, while the angular dependence is naturally imposed by the geometry of the pseudoplane region. Here we use the typical values $t_1 = 0.4$ eV and $t_2 = -0.08$ eV [55], and leave $\mu$ to adjust the doping concentration.

Along lines of $k_y = 1.00\pi$, $0.81\pi$, $0.62\pi$ and $0.50\pi$ in k-space, the calculated ARPES spectra, as shown in SI Fig. S8, agree well with previous ARPES experiments[29]. It is clear to see that a gap opens, with the maximum in the antinodal direction, i.e., $k_y = 1.00\pi$. When $k_y$ moves away from $1.00\pi$, the degeneracy is gradually lifted with the vanishing effect of $U_f$ and the recovery of Fermi surface finally. The antinodal gap formation can be attributed to the strong interaction between hole and charge pseudoplanes to form pQon, where the more localized p-state behavior is the most prominent and its interaction with the more delocalized Q-state is the strongest in the axial direction. We want to emphasize that our simulations have captured several unusual features in the spectra of the pseudogap phase[29]. First, the energy maximum of the antinodal gap (the band bending point at k$_G$ in Fig. 5a) is not at the Fermi momentum, $k_F$, where the gap due to BCS pairing locates. In the pseudogap phase, it was reported that the bending point of the dispersion relation is located at $k = k_G > k_F$[29,37]. This unusual feature of the electronic structure of the pseudogap is captured by the renormalized band structure of pQon, as shown in Fig. 5a, and the corresponding ARPES simulations, as shown in Fig. 5b and Supplementary Fig. S8. Second, the depth of the gap is almost doubled at $k_F = (0, \pi)$ than that at the $k_G$ point.

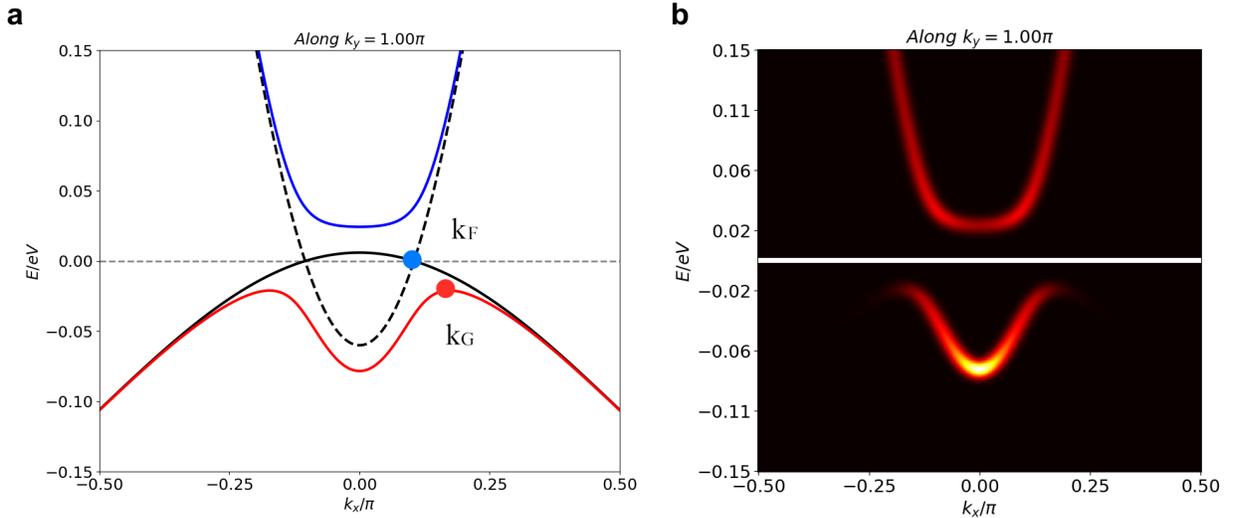

**Figure 5. Band structure and the calculated ARPES spectrum within the pQon model**. **(a)** Band structure from Eq. (3). **(b)** The corresponding calculated ARPES spectrum along $k_y = \pi$. Here we set $E_d = -0.1 E_k$.

We further add the temperature and doping dependences to $U_f$. In previous two-component models[56], the coupling of carriers to the lattice would decrease upon increasing temperature. Although Hall measurements show that the number of thermally activated holes increases exponentially versus temperature[57], suggesting that more itinerant carriers are released with increasing temperature, this is not the exact temperature dependence of interest here about the charge flux confinement capability of charge pseudoplanes, which directly modulates the duality interaction in pQon. Our DFT calculated charge leakage out of the pseudoplane region, as discussed in Fig. 2, in fact increases linearly with temperature, suggesting a linearly reduced duality interaction strength. We thus employ the $1/T$ form in Eq. (6). As to the doping dependence, since scanning tunneling microscopy (STM) measurements show that the energy scale of pseudogap rises linearly with decreasing doping[48], we make $U_f$ also decrease linearly versus doping, so that it terminates at a doping concentration $p = p_0$:

$$\boldsymbol{U_f} = V_f \frac{1}{T} \frac{p_0 - p}{p_0} |\cos k_x - \cos k_y| \tag{6}$$

Note that the expression is valid only when $p < p_0$. Equation (6) suggests that the pQon character and the unique electronic interaction strength of the pseudoplane region start to weaken with increasing temperature, while there is not an explicit characteristic temperature corresponding to the close of the pseudogap. Such a descriptor, as we will discuss later, is largely embedded in the $V_f$ and the doping terms instead. Besides, $U_f$ becomes zero when the doping concentration reaches $p_0$, and the carrier in cuprates will lose the pQon character, making the system behave more like a Fermi liquid[58]. Within the proposed picture, the phenomenon that the 'nodal' metal emerges in the pseudogap phase from the Fermi-liquid-like background with reducing doping[59], is driven by the increasing strength of $U_f$. However, the gap-like feature does not open immediately when $p < p_0$, instead, the spectrum weight first weakens around antinodal, with the pseudogap only opening up at a lower doping often below the optimal doping (Fig. 6a-c). That is also to say that when the pseudogap closes, $U_f$ is still a non-zero value. We ascribe this to the fact that the value of the antinodal gap is small enough to be overwhelmed by the lifetime broadening.

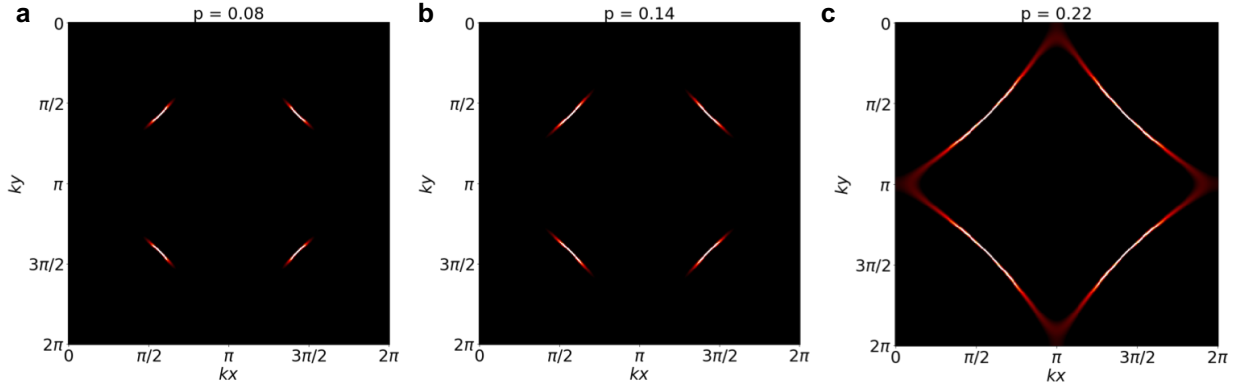

**Figure 6.** Calculated doping evolution of Fermi arc at $T = 50K$. **(a)** $p = 0.08$ **(b)** $p = 0.14$ **(c)** $p = 0.22$. Other parameters are $V_f = 6.28$ and $p_0 = 0.2659$, as obtained from Table 1 for Bi2212.

Another important property of the pseudogap phase is that the gapless region evolves from a Fermi arc of limited length to the non-interacting Fermi surface with increasing temperature[60]. We calculate the ARPES spectra at $\omega = 0$, as shown in Supplementary Fig. S9a-c, which show that with increasing temperature and decreasing $U_f$, as described by Eq. (6), the Fermi arc gradually expands, and eventually the non-interacting Fermi surface recovers, consistent with experimental observations[61]. The growth of the arc with temperature was explained by static disorders, such as charge density wave or spin density wave[24–26]; here we show that, instead, it could also be caused by the destruction of the strength measure of charge pseudoplanes, $U_f$, with increasing temperature. Besides, the arc is not a real locus of quasiparticle states, rather it forms because real spectral weight scatters into the d-wave-type gap at finite lifetime broadening $\Gamma$. Therefore, the difficulty in observing the back side of the arc is then natural in our picture, since the arc should be just arc-like in hole doped cuprates, instead of being the front side of a small pocket. Note also that we fix $t_1 = 0.4$ and $t_2 = -0.08$ and the $d$-wave form in the simulation, thus the curvature of the Fermi arcs is also fixed in Fig. 6 and Supplementary Fig. S9. In reality, both the hopping parameters and the d-wave form may be affected by temperature and doping, and hence the curvature of the Fermi arcs may change.

**Pseudogap and Superconducting Temperatures from Flux Confinement**

Significantly, when the gap equals the lifetime broadening $\Gamma$ in ARPES, i.e., $A(k, \omega = U_f) = A(k, \omega = 0)$, we have $U_f = \sqrt{2}\Gamma$ (See Methods). We thus obtain the temperature $T^*$ that marks the boundary of the pseudogap phase as a function of doping

$$T^* = \frac{V_f}{\sqrt{2}\Gamma} \frac{p_0 - p}{p_0} \qquad (7)$$

Note that $T^*$ in Eq. (7) depends on characteristic parameters of the specific experimental technique, which provides an explanation for the slightly different boundaries measured for the same material from various techniques. Eq. (7) can be well fitted to the experimental data of pseudogap temperature versus doping $T^*(p)$ [62–66]. The fitted result to Eq. (7) are shown in Supplementary Fig. S10, with the fitting parameters $V_f$ and $p_0$ summarized in Table 1. It is clear that the duality interaction strength $V_f$, or the intrinsic charge flux confinement property of the charge pseudoplane region, increases as the superconducting $T_{c,max}$ increases from LSCO to YBCO and to Bi and Hg families. Meanwhile, the pseudogap temperature $T^*$ is also proportional to $V_f$ in a wide doping range from 0 to $p_0$. Eq. (7) thus unveils that both pseudogap and superconducting phases are born from and intertwined through the charge flux confinement property of the charge pseudoplane region in cuprates. The increase of $T_c$ with the number of $CuO_2$ layers within a given family, as we argued in Ref. [18] and Fig. 2d, is however at least strongly correlated with the screening effect provided by increasing $CuO_2$ layers as the generalized pseudoplanes, to minimize the perturbation to the in-plane hole hopping by different apical mode symmetries, which does not obviously require the change of $V_f$ of the vertical pseudoplanes. We thus do not expect $V_f$ to reflect the exact layer dependence in general, although certain limited correlations can still emerge, such as in the Hg family (Fig. 2c). Based on Eq. (7), $p_0$ is the intersection of $T^*(p)$ on the dopoing axis as shown in Supplementary Fig. S10. Note that our ARPES simulations in Fig. 6 and Supplementary Fig. S9 used the paramters for Bi-2212 in Table 1. If using the parameters for LSCO, with smaller $V_f$ than Bi-2212, the simulated gap also closes at a lower doping, consistent with experimental observations[67–70].

| Materials | $T_{c,max}$ (K) | $V_f$ (eV · K) | $p_0$ |
|---|---|---|---|
| LSCO | 38 | 3.97 | 0.2575 |
| YBCO | 93 | 5.62 | 0.2835 |
| Bi2201 | 40 | 6.91 | 0.2409 |
| Bi2212 | 93 | 6.28 | 0.2659 |
| Hg1201 | 94 | 8.61 | 0.1801 |
| Hg1223 | 135 | 9.41 | 0.2468 |

**Table 1**. Fitted $V_f$ and $p_0$ according to Eq. (7) for LSCO [62], YBCO [63], Bi2201[64], Bi2212[65], Hg1201[66], and Hg1223[66]. A typical lifetime broadening of 10 meV is used.

**Temperature-invariant Critical Doping around 19% and STM inhomogeneity**

It was reported from Hall coefficient measurements that the normal state carrier density shows an abrupt increase at 19% hole doping of YBCO[71]. There is also a very recent report showing that the critical doping at around $p = 0.19$ in Bi-2212 is temperature-independent by ARPES measurements[45], where the strange metal abruptly reconstructs into a more conventional metal with Bogoliubov quasiparticles. We propose that such a temperature-independent effect is likely to have a geometric origin due to percolation. Using the pQon picture, we describe such critical doping as the transition point, beyond which the hopping process of pQons is forced to generate itinerant quasiparticles. This corresponds to when the pQon hopping pathways no longer form a percolated network due to the high pQon concentration. Figure 7 shows the result of our percolation simulation. Figure 7a to 7c are three examples of simulated lattice corresponding to doping concentration $p = 0.15, 0.19$ and $0.23$, respectively, where the black regions are formed by unavailable sites for a hopping pQon to land, either due to a direct pQon occupation or due to the fact that the open site is surrounded by other pQons too closely (see Methods and Fig. 7h). The white regions are the possible pathways where the pQons can hop and land, i.e., diffuse. Note that our picture here involves only one type of carrier of pQon, and the black/white regions and diffusion pathways are dynamically self-organized by the hopping induced redistribution of pQons. Although our picture is different from the two-carrier/component model[56] and/or defect-induced disorder or inhomogeneity[72], the defects can serve as the role to quench the intrinsic inhomogeneity discussed here.

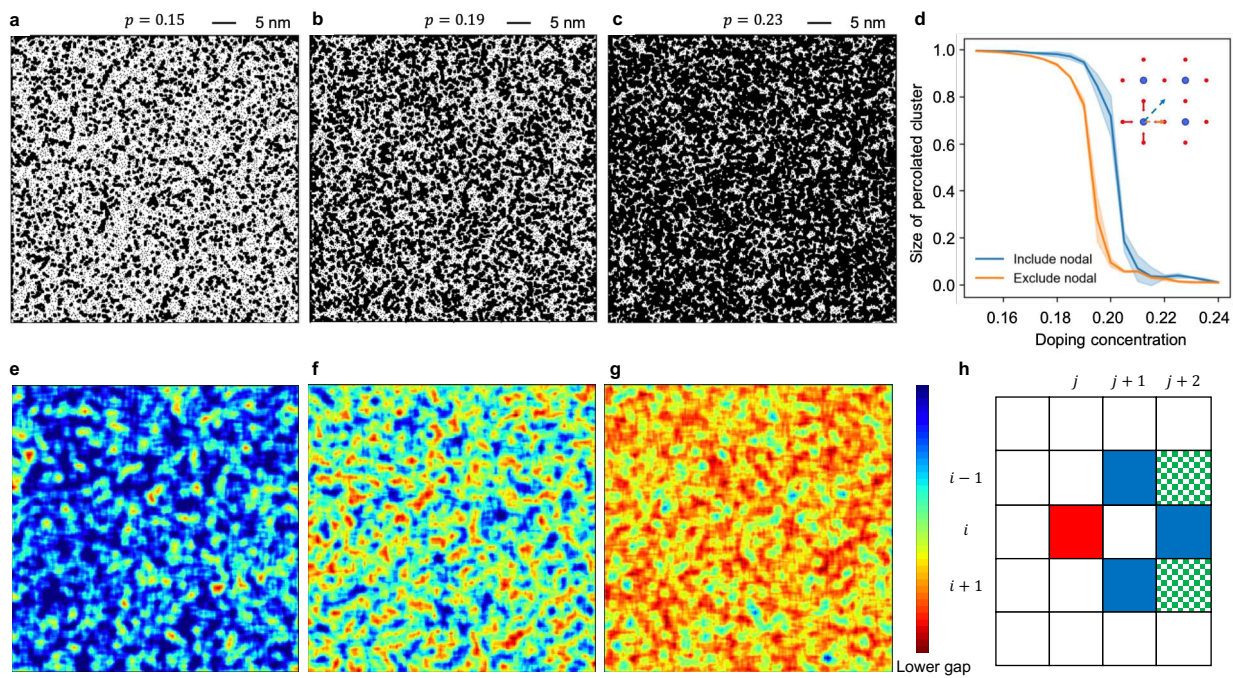

**Figure 7. Percolation of pQon hopping pathways and critical hole doping concentration. (a-c)** Examples of simulated 300*300 square lattice with doping concentration $p = 0.15$ (a), $p = 0.19$ (b), and $p = 0.23$ (c). The black regions correspond to sites unavailable to the landing of a hopping pQon. The white regions are formed by sites that are available to such landing. **(d)** Percolated cluster size defined as the size of the largest connected white cluster over the total size of white regions in the simulation. Shaded region represents the standard deviation as a result of random sampling. **(e-g)** Simulated pseudocolor images corresponding to $p = 0.15, 0.19$ and $0.23$. More red color indicates lower gaps. **(h)** Schematic illustration of the criteria used in the simulation. For a pQon at $(i, j)$ (red site), if any of the three blue sites is occupied by another pQon or both the green sites are occupied, the pQon at the red site is not allowed to hop toward the rightward quadrant. Same rule applies to other three quadrants. See Methods for further details.

The percolation transition is shown in Fig. 7d, where two schemes of 1) including both the nodal and antinodal hoppings and 2) just including the antinodal hopping are compared (see Methods). They actually give similar percolation thresholds around 20% and 19%, respectively. Thus, the examples in Fig. 7a-c correspond to the percolated, the threshold, and the non-percolated pQon hopping pathways, respectively. This suggests that our method provides a percolation threshold robust against minor variation of the settings in the model, which is likely to be responsible for the measured critical doping around 19%. If the pQon diffusion pathways form a percolated network ($p < 19\%$), pQons are well confined during the hopping processes. While beyond 19%, the landing event in a pQon hopping process will be frequently impeded by the unavailable sites due to the lack of a percolated pathway of available sites, leading to an itinerant quasiparticle (*not* the more delocalized Q-state of pQon) observed in ARPES and Hall measurements. However, these Bogoliubov quasiparticles are in a dynamical conversion with pQons, that is, after certain lifetime they will be converted back to pQons to maintain a constant ratio between the two types of carriers at a given hole doping level beyond 19%.

Furthermore, the simulation suggests a cluster size of around 5 nm, which corresponds well with the scales of the spatial variation of gap sizes in previous STM images[46–48]. We then simulate the pseudocolor images in Fig. 7e-g based on our results in Fig. 7a-c by assuming that a pQon surrounded by more unavailable neighboring sites locally is easier to be converted to an electric signal in the STM measurement, corresponding to a smaller local gap. These pseudocolor images compare well with previous STM images showing spatial inhomogeneities in local gap sizes. The results thus suggest that the STM tip may serve as a local probe with the bias voltage to extract the information about the spatial variation in pQon hoppings, which is often averaged out in ARPES measurements and only emerges as a collective behavior beyond 19%. In the case of STM, instead, the local bias voltage forces the pQon in the black and white regions to respond differently in a broad doping range. Note that due to the different types of perturbations provided by the local bias voltage in STM and the global photon excitation in ARPES, and different methods in extracting the gap values in STM measurements and analysis, a percolation threshold obtained from analyzing STM images may slightly deviate from 19%. In summary, our percolation simulations based on pQons show that both the STM inhomogeneity and the 19% critical doping in ARPES may originate from the duality nature of pQon and their collective distribution and interaction in the $CuO_2$ plane.

In conclusion, we have identified the charge pseudoplane region as a critical emergent configuration that could play a central role in understanding the complicated phenomena in cuprate superconductors. The pseudoplanes cut through the $dx^2$-$y^2$ and $dz^2$ bonding directions and possess a material-dependent, dynamic charge- flux confinement property. We have further proposed pQon as the new fundamental collective excitation in cuprates, with momentum-dependent and ultrafast localization-delocalization duality. Our ARPES simulations based on the electronic structure of pQon achieve a good agreement with experimental ARPES measurements in a broad doping range. Our picture also naturally leads to a geometric percolation interpretation of the temperature independent critical doping $p_c = 0.19$ reported recently by ARPES, as well as previous STM imaging results on spatial inhomogeneity, all from the collective hopping interaction of pQons. Our work therefore provides a new perspective to reconcile the results from various experimental techniques with different momentum, spatial and time resolutions. More importantly, we find that the duality interaction strength in a pQon that reflects the flux confinement ability of pseudoplanes, among different cuprate families decides $T_{c,max}$, which could serve as a new descriptor for the design and search of higher temperature superconductors. We believe that our picture of pseudogap phase related to the collective confinement behavior of charge fluxes will help elucidate the origin of pseudogap phase and other related phases in high temperature superconductors.

**Methods**

**DFT simulations**

All DFT calculations were carried out by the Vienna Ab initio Simulation Package (VASP) code, which implements the pseudopotential plane wave band method[73]. The projector augmented wave Perdew-Burke-Ernzerhof (PAW-PBE) functional was utilized for the exchange-correlation energy. A 520 eV plane-wave energy cutoff was implemented for all calculations. Non-spin polarized DFT+U calculations with U = 8 eV and J = 1.34 eV[74] integrated with PHONOPY package [75] were used for phonon simulations using the supercell approach with finite displacements based on the Hellmann-Feynman theorem. The phonon modes are enumerated in a 2x2x1 supercell for all Hg-1212, YBCO, and LSCO. 21 representative phonon modes out of the total 36 modes of Hg-1212 were selected for the calculation of phonon couplings. The definition of the 2x2x1 cell is shown in Fig. 1b, 1c. 16 out 24 modes of the Hg-1201 supercell and 24 out of 39 modes of the YBCO supercell with the same size are selected. The other modes have similar ion movements to the selected modes and were excluded due to the computational limit. We expect that adding those similar modes back will not induce any new trend in Fig. 2. Because of the smaller number of atoms and more symmetry elements of LSCO, all the 21 modes are selected. The phonon modes were calculated at $k = (0.5, 0.5, 0)$. For each pair of phonons, one is chosen as the "frozen mode" and one as the "perturbation mode". The ionic distortions corresponding to the frozen mode is added to the structure. Five calculations were performed on each frozen mode at a given amplitude, with the added amplitude from each perturbation mode scaled at -0.5, -0.25, 0, 0.25, 0.5 of the frozen mode amplitude. The five energies of each such calculation is used to fit a parabola energy profile to extract the strength of phonon coupling by the energy of the equilibrium position of the fitted curve, following previous method[18]. If the two modes have no anharmonic coupling, the minimum of this energy profile will be at the zero amplitude of the perturbation mode. If there is an anharmonic coupling, the minimum will be shifted to a nonzero amplitude. Suppose the frozen mode is mode $x$ with 1.0 amplitude and the perturbation mode is mode $y$ with 0.25 amplitude, we first take the calculated 3D charge density (CD) with both modes applied and subtract the charge density from only the frozen mode $x$ and 0.25 of the charge density from only the perturbation mode $y$, i.e., CD($x+y$) – CD($x$) – 0.25*CD($y$), which gives the total extra charge flux induced by the pair of phonon modes. Among these extra charge fluxes, the ones outside the pseudoplane region is the value of extra flux leakage in positive or negative number of electrons. The ratio between the latter and the former is defined as the extra flux leakage in Fig. 2b (i.e., a ratio without unit). In Fig. 2d, however, we only consider the extra charge flux in a 1 Å thick slab region in the *ab* plane centered around the $CuO_2$ plane as the generalized pseudoplane, since the $CuO_2$ plane is directly and strongly affected by the apical mode symmetry. In the illustration of the flux flow pattern in Fig. 1c, only the four oxygens around one Cu in the $CuO_2$ plane and the apical oxygen above this Cu are displaced according to the breathing mode and apical mode. The strength and direction of the charge fluxes are obtained from the charge density difference between the distorted and original structures. For example, Supplementary Fig. S1 shows the cross sections of charge density difference at different *c* values. In Fig. S3, all the hypothetical structures were relaxed, and the same atomic displacements of each mode as in Fig. 2a were used. For the localization-delocalization conversion calculation in Fig. 4b, Heyd–Scuseria–Ernzerhof (HSE) exchange-correlation functional with $\alpha = 0.25$ and a 2x2x1 supercell was used. The localized hole was created by HSE relaxation following previous method[18]. The quasiparticle was created by HSE relaxation with symmetry enforced as well as non-perturbed initial ionic positions in a charged supercell. The intermediate images and energies in Fig. 4b were generated by linear interpolation of the ionic positions in initial and final images followed by electronic self-consistent relaxations.

**ARPES simulations**

The spectral functional was calculated using the (1,1) component of Green function within our pQon model:

$$G^{-1}(k, \omega) = \omega + i\Gamma - \varepsilon_{Q,k} - \frac{U_f(k)^2}{\omega + i\Gamma + \varepsilon_{d,k}}$$

where $\omega$ is the energy, and $\Gamma$ represents lifetime broadening (set as a constant of 10 meV in our simulations). Note that both higher energy $\omega_{k\sigma}^+$ and lower energy $\omega_{k\sigma}^-$ bands are involved as poles of the Green's function. The spectral weight is defined by:

$$A(k,\omega) = -\frac{1}{\pi}Im(G(k,\omega))$$

For Fig. 5, $k_y$ was fixed when plotting the spectral weight versus $k_x$ and $\omega$. For Fig. 6, $\omega$ was fixed at zero and the spectral weight was plotted versus both $k_x$ and $k_y$ in the entire Brillouin zone.

**Percolation related simulations**

A 300*300 square lattice model was used for the percolation simulation to model the copper sites in one CuO$_2$ plane. Each grid of the lattice corresponds to a copper site. The hole (i.e., pQon) distribution corresponding to a doping concentration $p$ was generated by random sampling while avoiding nearest-neighboring holes. A sampled grid with any occupied nearest neighbors was discarded and resampled, until a grid with no occupied nearest neighbors was obtained. After obtaining a legal sampling, all hole positions were first marked black. For a hole currently sitting at site $(i,j)$, its ability to hop in the model is given by the following criteria (Fig. 7h). When considering four quadrants $\{(-\frac{\pi}{4},\frac{\pi}{4}),(\frac{\pi}{4},\frac{3\pi}{4}),(\frac{3\pi}{4},\frac{5\pi}{4}),(\frac{5\pi}{4},\frac{7\pi}{4})\}$ as its possible hopping directions, for the quadrant $(-\frac{\pi}{4},\frac{\pi}{4})$, if at least one site of $\{(i,j+2),(i+1,j+1),(i-1,j+1)\}$ was occupied, or both sites $\{(i+1,j+2),(i-1,j+2)\}$ were occupied by other holes, the hole at $(i,j)$ was considered as not movable to the $(-\frac{\pi}{4},\frac{\pi}{4})$ quadrant. The same rule was applied to the analysis of other quadrants to determine whether the hole can hop along other directions. If the hole could not move along at least three quadrants (Setting 3 rather than 4 quadrants as the criterion here assumes a strong hopping competition with other holds to the remaining quadrant), this hole was classified as not movable and its four nearest-neighbor sites and four second nearest-neighbor sites would be also marked as black, as shown in Fig. 7a-c. The ratio of the size of the largest connected white cluster over the size of all white grids is the indicator of percolation. The percolation threshold is defined as the doping concentration where such ratio swiftly changes from very close to zero to very close to one. Note that for infinitely large lattice (i.e., the thermodynamic limit) the transition should be a step function. For finite-size simulation the corners are slightly rounded as shown in Fig. 7. In this case, the midpoint of the step (ratio = 0.5) was used as the approximated percolation threshold. For pseudocolor image simulation (Fig. 7e-g), we assume that a pQon that is surrounded by more unavailable neighboring sites locally is easier to be converted to an electric signal in the STM measurement, thus having a smaller gap measured by STM. We thus let the black regions in the percolation simulation have gap value 0 and white regions have gap value 1, and average over every 8*8 (about 1.5Å*1.5Å) region to produce the pseudocolors in Fig. 7e-g.

**Author Contributions**

X. C. and J. D. contributed equally to this work. X. L. conceived the physical picture, designed and supervised the research. X. C. performed the DFT and percolation simulations. X. L. and J. D. wrote the Hamiltonian. J. D. performed the ARPES simulation. All authors analyzed the results and wrote the manuscript.

1–5 (2015).

**Supplementary Information**

# A picture of pseudogap phase related to charge fluxes

Xi Chen[1,†], Jiahao Dong[1,†], Xin Li[1,*]

1. John A. Paulson School of Engineering and Applied Sciences, Harvard University

29 Oxford St, Cambridge, MA 02138

*: Corresponding author: lixin@seas.harvard.edu

†: Equal contribution

**Charge flux and electronic structure calculations**

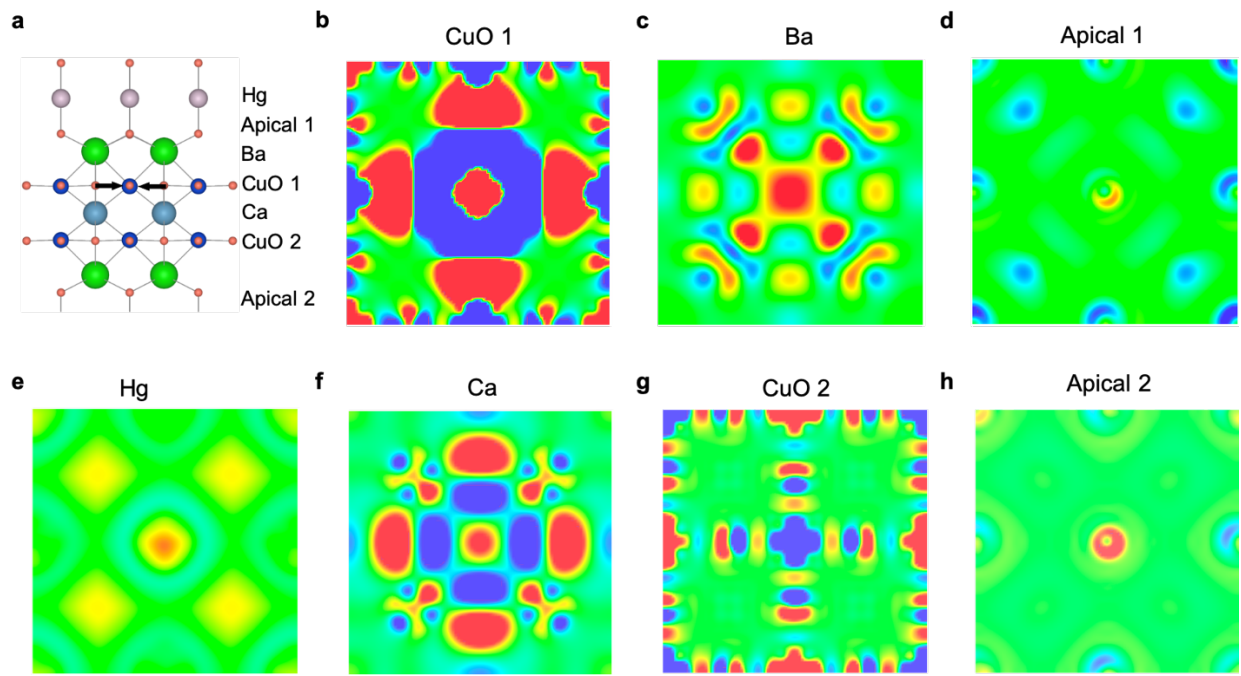

Figure S1. Charge density difference induced by the breathing mode. (a) Illustration of a breathing mode and the label of each cross-section perpendicular to the *c* direction shown in (b) – (h). (b) – (h) Cross sections of the charge density difference induced by the breathing mode. The more red (blue)-ish regions correspond to increased (decreased) charge density.

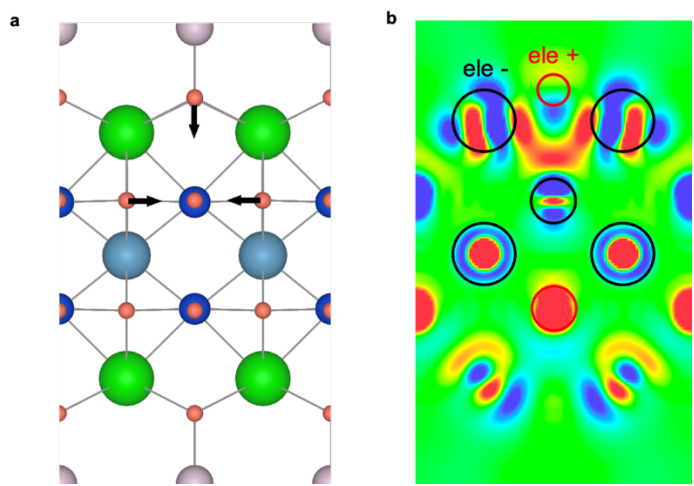

Figure S2. Reduction of out-of-pseudoplane charge flux when breathing mode couples to apical mode. (a) Illustration of the coupled modes by black arrows. (b) The corresponding cross section of the charge density difference between the case of breathing mode coupling to the apical mode, and the breathing mode alone. The more red (blue)-ish regions correspond to increased (decreased) charge density. Black (red) circles indicate that the total charge inside is decreased (increased).

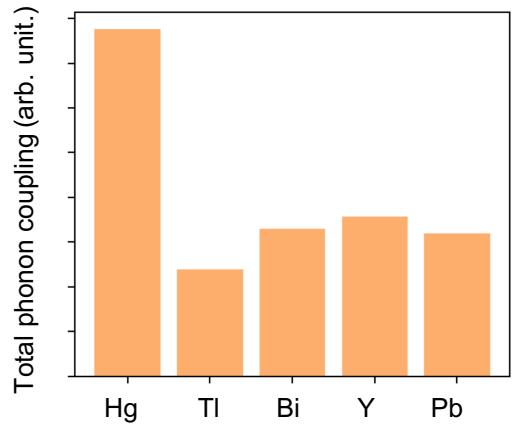

Figure S3. Total phonon coupling of Hg-1212 (Hg) and that of structures by replacing the apical atom Hg with Tl, Bi, Y and Pb. The total phonon coupling is the sum over the bottom three rows of the Hg-1212 case in Fig. 2a that includes apical, breathing and anti-breathing modes (frozen mode #19-21), respectively. They also couple with many perturbation modes to prevent the charge flux leakage in the same bottom three rows in the Hg-1212 case in Fig. 2b.

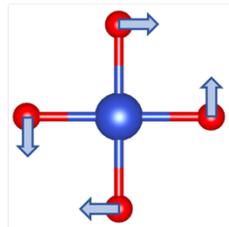

Figure S4. Illustration of the rotation oxygen mode.

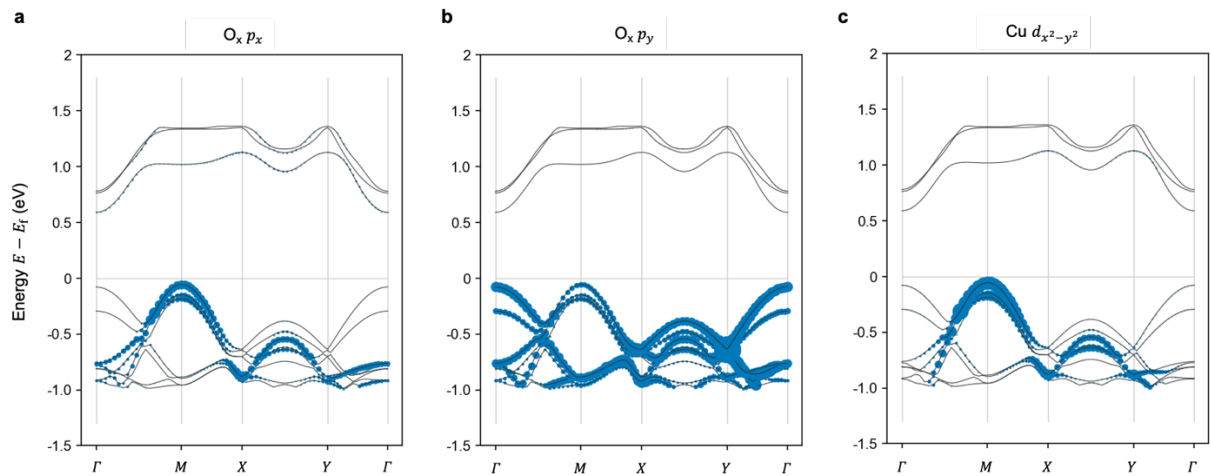

Figure S5. Band occupation of the $O_x$ oxygen $p_x$ (a) and $p_y$ (b) orbitals and the copper $d_{x^2-y^2}$ orbital when an oxygen rotation mode (Fig. S4) is applied together with the apical mode that does not prefer to couple with each other.

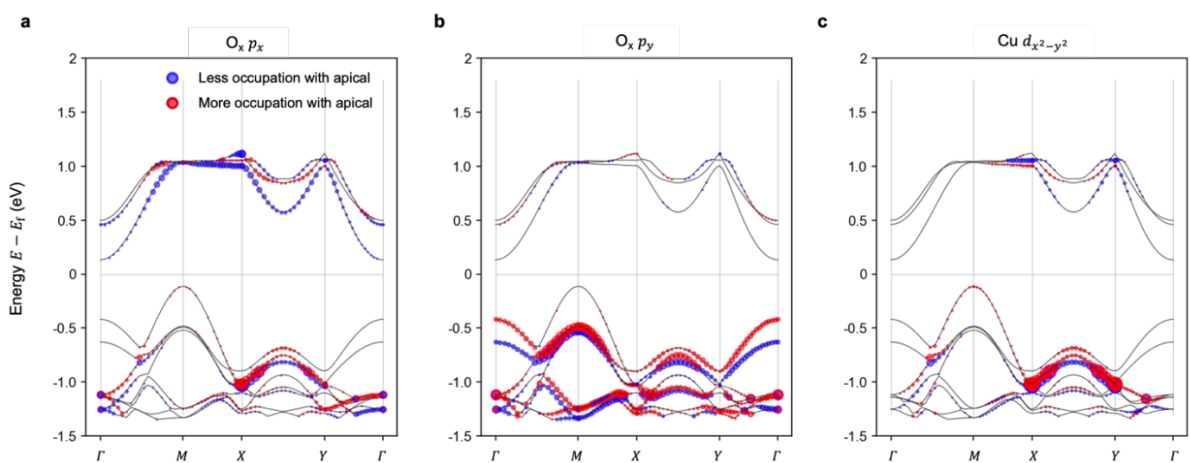

Figure S6. Change in the band occupation of the $O_x$ oxygen $p_x$ (a) and $p_y$ (b) orbitals and the copper $d_{x^2-y^2}$ orbital when the breathing mode is coupled to the apical mode. Red and blue spheres correspond to more and less occupation with coupling to the apical mode. The referenced band structure is the one with breathing mode and apical mode coupling.

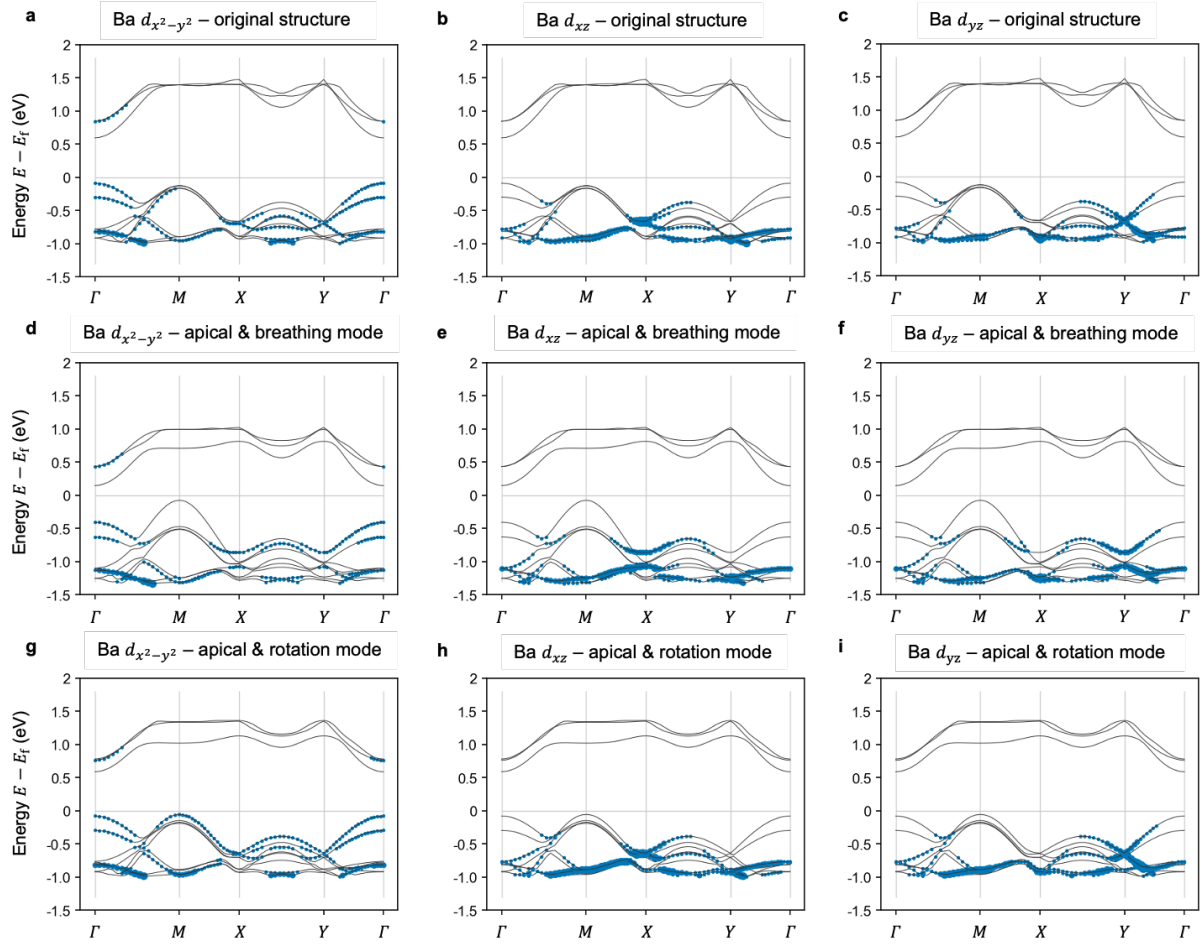

Figure S7. Band occupation of the three nonempty Ba orbitals.

**Additional ARPES simulations**

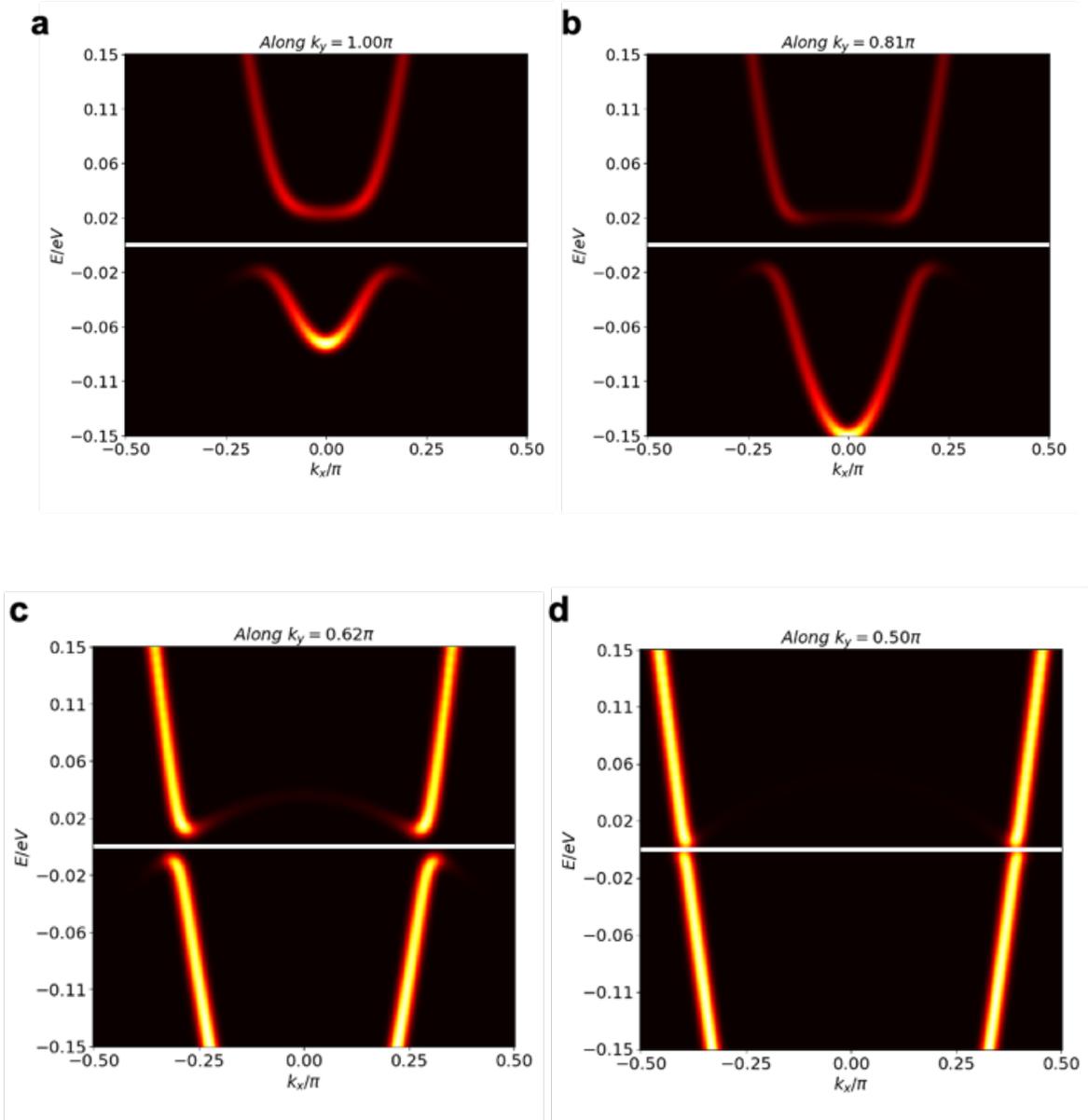

Figure S8. Calculated ARPES spectral weight within the pQon model, along (a) $k_y = 1.00\pi$, (b) $0.81\pi$, (c) $0.62\pi$ and (d) $0.50\pi$.

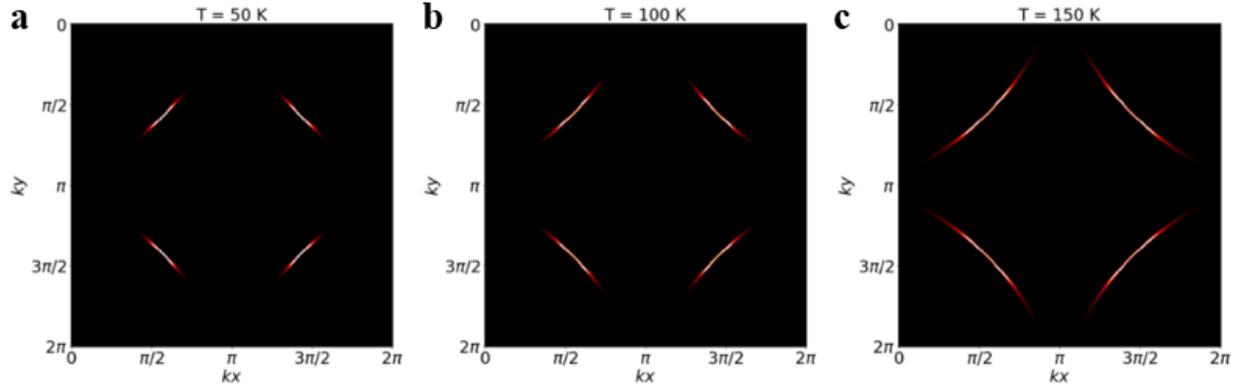

Figure S9. Calculated temperature evolution of Fermi arc at $p = 0.14$ for (a) $T = 50K$, (b) $T = 100K$ and (c) $T = 150K$. Other parameters are $V_f = 6.28$ and $p_0 = 0.2659$ as obtained from Table 1 for Bi2212.

## $V_f$ and $p_0$ obtained from $T^*(p)$ relationship

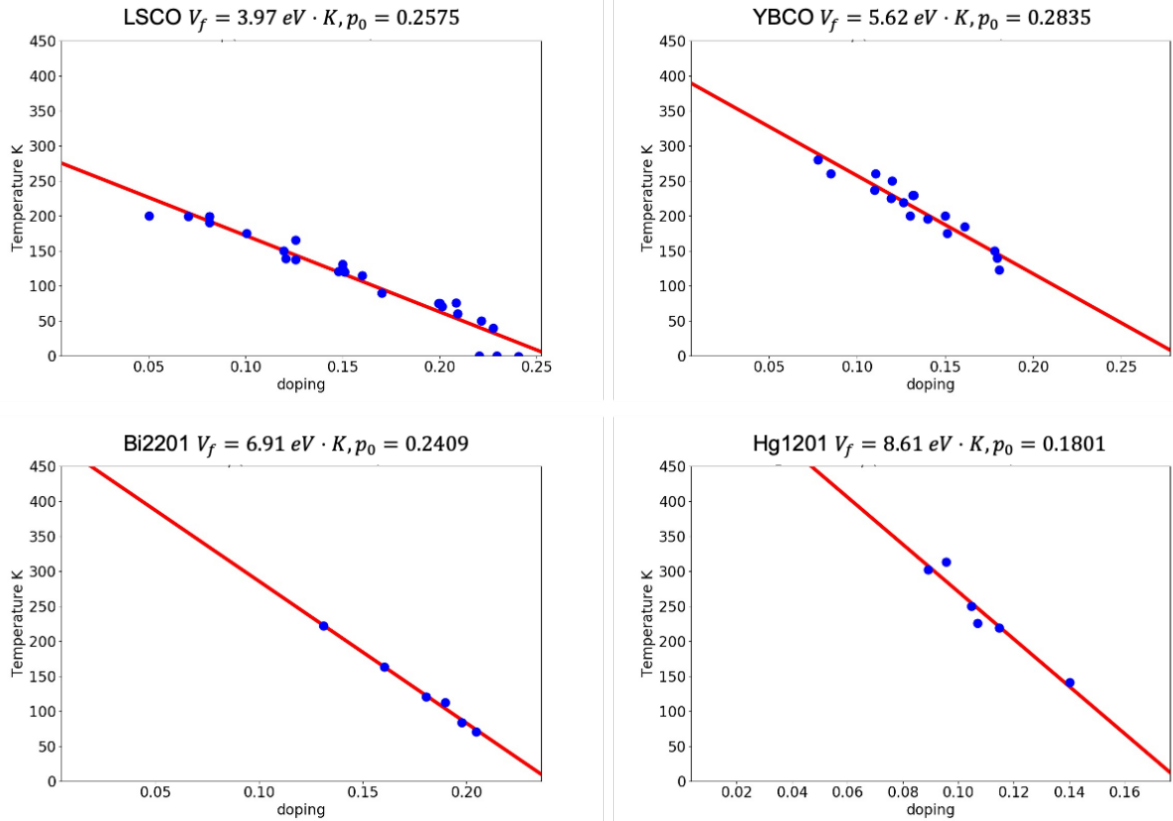

Figure S10. $V_f$ and $p_0$ obtained by fitting $T^*(p)$ derived in Eq. (7) to the experimental boundary of pseudogap phase versus doping. Blue dots are experimentally measured boundary of pseudogap phase and red lines are fitted according to Eq. (7).